\documentclass[twocolumn]{aa}
\usepackage{natbib}
\usepackage{color}
\usepackage{ragged2e}
\usepackage[pdfpagelabels=false]{hyperref}	
\hypersetup{colorlinks=true,linkcolor=blue,citecolor=blue,filecolor=blue,urlcolor=blue,}
\usepackage[varg]{txfonts}
\usepackage{graphicx,rotating}
\usepackage[normalem]{ulem}
\usepackage{colortbl}
\usepackage{xcolor}
\usepackage{bbold}

\newcommand{\msun}{\rm M_\odot}

\bibpunct{(}{)}{;}{a}{}{,} 

\definecolor{darkolivegreen}{rgb}{0.33, 0.42, 0.18}
\definecolor{salmon}{rgb}{0.95,0.5,0.25}

\begin{document}

\title{Distinguish dark matter theories with the cosmic web and next-generation surveys I: an alternative theory of gravity}

\titlerunning{Cosmic web in an alternative theory of gravity}
\author{Pierre Boldrini \inst{1} \and Clotilde Laigle \inst{1}}

\offprints{Pierre Boldrini, \email{boldrini@iap.fr}}
\institute{$^{1}$ Sorbonne Universit\'e, CNRS, UMR 7095, Institut d'Astrophysique de Paris, 98 bis bd Arago, 75014 Paris, France}

\authorrunning{Boldrini et al.}
\date{Accepted}

\abstract{In the context of future large surveys like the Euclid mission, extracting the cosmic web from galaxies at higher redshifts with more statistical power will become feasible, particularly within the group-cluster mass regime. Therefore, it is imperative to enlarge the number of metrics that can be used to constrain our cosmological models at these large scales. The number of cosmic filaments surrounding galaxies, groups and clusters, namely the connectivity, has recently emerged as a compelling probe of the large-scale structures, and has been investigated in various observational and numerical analyses. In this first paper, we examine dark matter-only cosmological simulations using the widely used {\sc DisPerSE} filament finder code under two theories of gravity: the Poisson ($\Lambda$CDM) and the Monge-Ampère models, in order to quantify how alternative models of gravity alter the properties of the cosmic skeleton. We specifically focused on this alternative gravity theory due to its propensity to enhance the formation of anisotropic structures such as filaments, but it also makes them more resistant to collapse, which consequently reduces the formation of halos. Indeed, our findings reveal that replacing the Poisson equation has a significant impact on the hierarchical formation scenario. This is evidenced by examining the redshift evolution of both the slope and the offset of the connectivity. Additionally, we demonstrated that current observations are generally in better agreement with our well-established gravity model. Finally, our study suggests that filament connectivity in the group-cluster regime could serve as a probe of our gravity model at cosmological scales. We also emphasize that our approach could be extended to alternative theories of dark matter, such as warm or fuzzy dark matter, given the extraordinary datasets provided by next-generation surveys.}

\maketitle

\begin{table*}
\begin{center}
\label{tab:landscape}
\caption{Connectivity in the literature}
\begin{tabular}{cccccccccccc}
 \hline
Cosmological & Box & $m_{\mathrm{DM}}$ & Cosmic web  & Persistence & Calibration \\
simulations & [Mpc/h] & [$10^{9}\msun$]&   tracers  &     $\sigma$   & method \\ 
    \hline
Poisson DM-only (This work) & 205  & 8  & DM / halos  & 7 / 1 & CP$_{\rm max}$-halos\\
Monge-Ampère DM-only (This work) & 205 & 8  & DM / halos  & 9.5 / 7  & CP$_{\rm max}$-halos\\
DM-only \citep{2010MNRAS.408.2163A} &  150 &  2 & DM   & - & -\\
HzAGN 2D \citep{2019MNRAS.489.5695D} & 100  & 0.08  & 2D galaxies  & 1.5  & HzAGN 3D\\
ITNG-300 \citep{Gouin2021} & 205  & 0.06  & halos & 3 & CP$_{\rm max}$-halos\\
The300 \citep{2023arXiv231007494S} & 1000  & 1.8  & gas & 0.2$^{1}$ & -\\
HzAGN 3D \citep{2020MNRAS.491.4294K}$^{*}$ & 100  & 0.08  & 3D galaxies  & 3  & -\\
MTNG \citep{2023arXiv230908659G}$^{*}$ & 7500 &  11 & 3D galaxies & 2 & CP$_{\rm max}$-halos\\
    \hline
    \hline 
 Observations  &  Redshift & Persistence  & Calibration & Number of clusters & Sky area\\ &  & $\sigma$ & method &  or groups  & deg$^2$\\
    \hline
COSMOS2015 \citep{2019MNRAS.489.5695D} & 0.5 - 1.2 & 1.5 & HzAGN 2D & 900 & 1.38\\
CFHTLS \citep{Sarron2019}   & 0.15 - 0.7 & 2 & 2D CFHTLS & - & 154\\
Coma \citep{Mala2020} & local & 5 & - & 1 &-\\
Abell 2142 \citep{Einasto2020} & local & - & - & 1 & -\\
SDSS DR10 \citep{Einasto21} & 0.009 - 0.2 & - & - & 4 & 112\\
    \hline
    \hline
\end{tabular}
\parbox{\hsize}{Notes: From left to right, the columns give: data type, the size of the box, the resolution and the cosmic web tracers for simulations, the redshift of observed data, the persistence threshold to detect filaments, and the calibration method to set this free parameter of the \texttt{DisPerSE} code. If the persistence is not specified, it means that the skeleton has not been extracted with the \texttt{DisPerSE} code. One of the calibration methods consists of comparing the positions of peaks (CP$_{\rm max}$) to those of the most massive halos such as galaxy groups and clusters ($> 10^{13} \msun$) (see section \ref{calib}). Concerning cosmological simulations, all the cosmic web tracers are 3D quantities except for HzAGN simulation. For COSMOS and CFHTLS data, the calibration method involves optimally recovering the filaments in the 2D light-cone catalogues created from 3D cosmological simulations. $^{1}$ It corresponds to an absolute persistence cut because \texttt{DisPerSE} is not applied to discrete particles but on a 3d grid for the gas. $^{*}$ These data are not taken into account in our study (for the reasons, see~\ref{DCon}).}
\label{tab1}
\end{center}
\end{table*}

\section{Introduction}

While its existence is inferred from its gravitational effects on baryons, the true nature of dark matter (DM) remains one of the most significant unsolved mysteries in astrophysics and cosmology. As it continues to encounter challenges on galactic and cosmological scales such as the cusp-core problem \citep{2017ARA&A..55..343B,2021Galax..10....5B}) or the $\sigma_8$ tension \citep{2019arXiv190105289D}, the standard cosmological model, $\Lambda$CDM, remains in competition with a variety of alternative DM models. A first possibility is that the DM is more complex and hotter than simple CDM. A wide range of alternative DM models has been proposed over the last decades to resolve $\Lambda$CDM issues. Three DM theories have gained attention as alternatives to the standard CDM model: warm dark matter \citep[WDM,][]{2001ApJ...556...93B}, self-interacting dark matter \citep[SIDM,][]{1992ApJ...398...43C,2000PhRvL..84.3760S}, and fuzzy dark matter \citep[FDM,][]{2000PhRvL..85.1158H}. They offer potential solutions to some of CDM puzzles, particularly on small scales. It is important to note that the traditional approach has always been to add new properties to DM to address the issues of the CDM paradigm. 

Rather than modifying the nature of DM, one could question whether Poisson's gravity provides a complete description of the gravity that DM undergoes on a cosmological scale. Even if the Poisson equation describes the large-scale structures of the Universe and agrees remarkably well with observations, alternatives to Poisson's gravity can be considered to explain the $\Lambda$CDM problems. Here, we focus on a specific alternative theory of gravity, Monge-Ampère gravity. Replacing Poisson's gravity with Monge-Ampère's simply involves changing the trace operation in the Poisson equation to a determinant operation as: 
\begin{equation}
    \mathrm{Tr}(\mathbb{1} + \frac{1}{4\pi G \bar{\rho}} \; D^2 \phi) =  \frac{\rho}{\bar{\rho}} \longrightarrow \mathrm{Det}(\mathbb{1} + \frac{1}{4\pi G \bar{\rho}} \; D^2 \phi) =  \frac{\rho}{\bar{\rho}}. 
\end{equation}
For weak gravitational potential $\phi$, by expanding the determinant about the identity matrix, one recovers the Poisson model approximately and exactly in one dimension. \cite{YB_MAG_2011,ambrosio2020mongeampere} proposed that the Monge-Ampère equation might provide an alternative model of self-gravitating systems. \cite{brenier:hal-01137528} has also shown that Monge-Ampère gravity can arise from a microscopic system in which a finite number of indistinguishable particles move on independent Brownian trajectories without any interaction. Then, gravity emerges through application of a principle of statistical physics, namely the large deviation principle. Monge-Ampère gravity has also been previously studied in the framework of Galileon gravity \citep{2000PhLB..485..208D,2003JHEP...09..029L,2009PhRvD..79f4036N,2009PhRvD..80f4015D} and could be a natural low-energy limit of the Galileon theory, similar to the Poisson equation in general relativity. Moreover, it is quite common in mathematics to replace Poisson with Monge-Ampère because it allows solving this nonlinear partial differential equation by approaching it as an optimal transport problem that can be solved with highly efficient algorithms \citep{Merigot11,Jun2019}. In other fields, in particular meteorology, invoking Monge-Ampère equation instead of Poisson equation was a fruitful approach: the semi-geostrophic model works far better on large scales than the older quasi-geostrophic model which instead uses the Poisson equation \citep{cullenBook}.

One of the challenges confronting alternative theories is determining effective theoretical experiments within a cosmological context. One major problem is that the contribution of baryons in a gravitational and hydrodynamical fashion on the distribution of DM within galaxies is non-negligible. Besides, the presence of baryons can potentially bias our constraints of the DM properties at small scales \citep[e.g.][amongst others]{gouin19,schneider19,chisari19}. Instead of trying to control for these biases, an alternative is to look at larger scales where gravitational dynamics is dominated by DM to test and confront alternative theories with observations. At these scales, DM is distributed along the anisotropic cosmic web \citep{Bond96, Dmitry}, a well-understood consequence of the gravitational growth of primordial overdensities \citep{zeldovich1970}. It forms a complex network made of filaments and walls bordering huge regions of low density, called voids, and intersecting at clusters of galaxies \citep[e.g.][]{Davis1982,Lapperent1986,Geller1989}. This large-scale cosmic web contains the diffuse material necessary to grow the virialized structures, up to 40$\%$ of the total matter in the Universe. The topology of the large-scale density field provides an interesting probe to constrain gravity or cosmology, in complement to standard statistics of galaxies and clusters, like the cluster mass function, the two-point correlation functions, or even higher order moments. This avenue has already been explored e.g. through Minkowski functionals \citep[e.g.][]{Gott86,2010ApJ...715L.185P,Hikage03,Appleby18}, statistics of voids \citep[e.g.][]{Sahlen16,Hamaus16,Contarini24}, the clustering of critical points of the density fields \citep[e.g.][]{shim23} or the properties of the skeleton \citep[e.g.][]{HoGronke2018} amongst other statistics.

In this paper, we propose an approach to quantify how an alternative model of gravity alters the properties of the cosmic network as measured through the connectivity, i.e. the number of cosmic filaments branching out a cluster. Actually, we measure multiplicity and not connectivity, in the wording of \cite{codis18}. Connectivity is a tracer of the growth of structures and of the percolation properties of the cosmic web \citep{pichon10} and as such its evolution throughout cosmic times is expected to informs us both about the equation of state of dark energy and the model of gravity which drives the coalescence of structures \citep{codis18}. Connectivity has recently emerged as an interesting probe of the large-scale structure around galaxies, groups and clusters, and was explored in several observational and numerical analyses. Pinpointing clusters, the most massive structures of the Universe, will ultimately mitigate the degeneracy between baryonic physics and the model of gravity. However this requires to observe sufficiently large area to beat the statistical uncertainties, and to overcome the fact that connectivity might be dependent on the degree of relaxation of the clusters \citep[e.g.][]{Gouin2021}. We expect that having a relatively large sample size will help minimizing the risk of being contaminated by systematic errors of such nature, or at least controlling their impact.

\begin{figure}[!]
\centering
\includegraphics[width=\hsize]{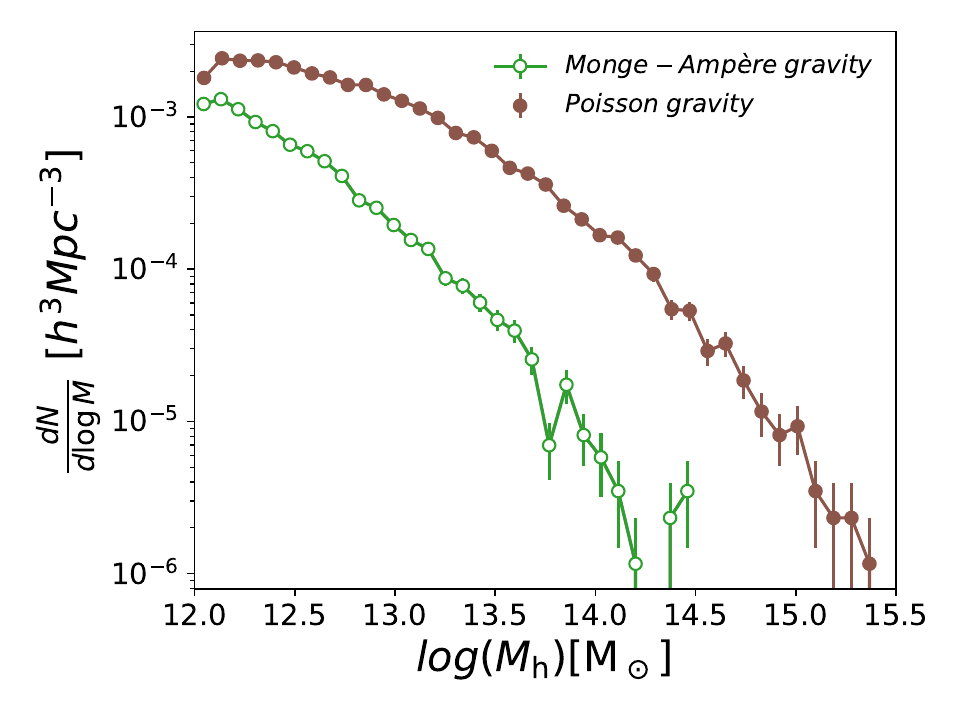}
\caption{Halo mass function of Poisson and Monge-Ampère cosmological simulations with $L=205$ Mpc/h and $512^3$ particles at $z=0$, as determined using the \texttt{adaptaHOP} halo finder \citep{2004MNRAS.352..376A}. Only (sub)halos that have more than 100 or more particles are kept. We have also included a Poissonian error normalized by the cosmological box size.}
\label{HMF}
\end{figure}

\begin{figure}[!b]
\centering
\includegraphics[width=\hsize]{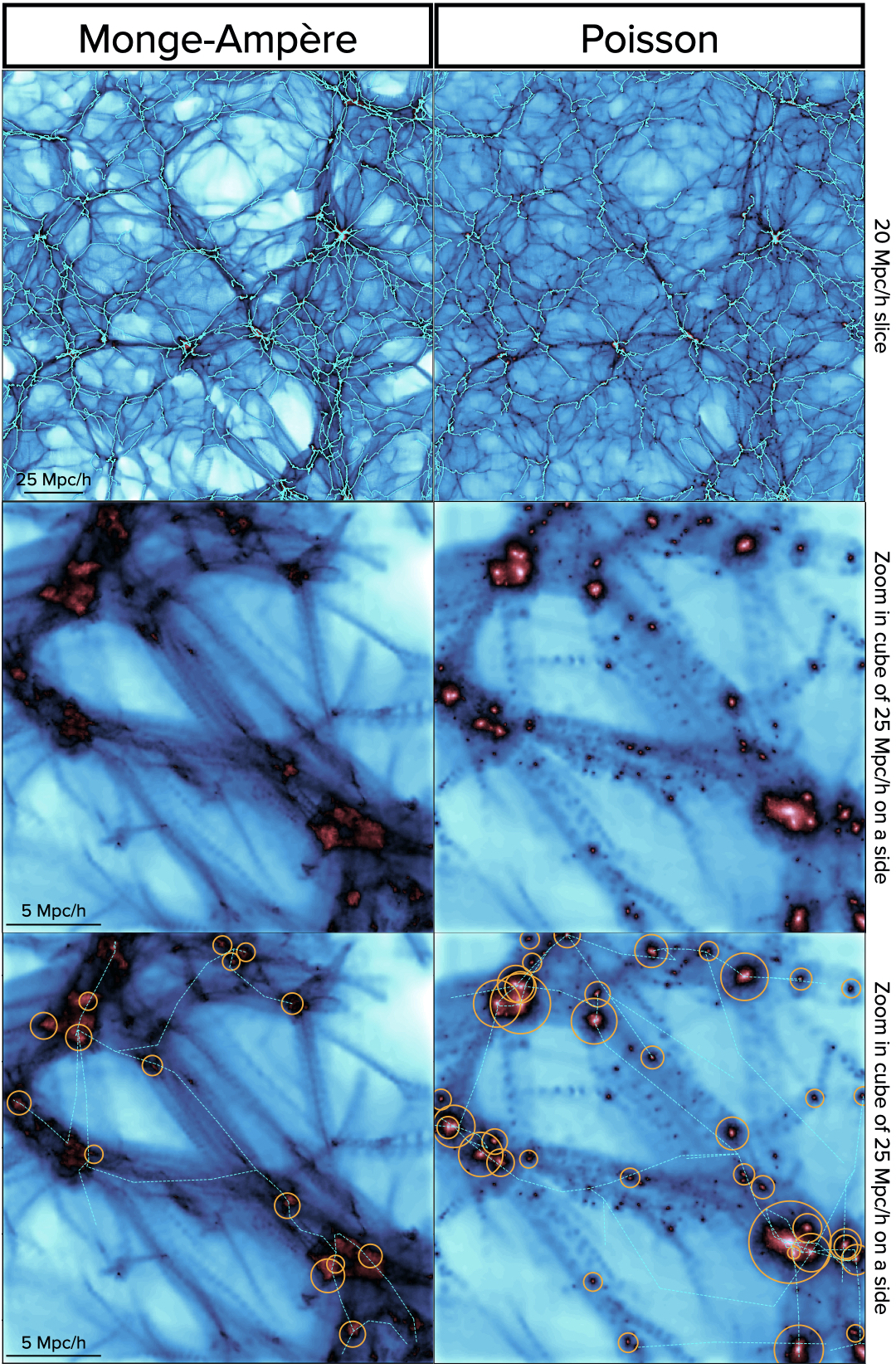}
\caption{Diversity of dark matter structures for Poisson and Monge-Ampère simulations. We show projected density maps of a 20 Mpc/h slice of the full cosmological box (top panels) and of zoomed regions of 25 Mpc/h on a side (middle and bottom panels). On top panels, we over-plotted the filaments identified by \texttt{DisPerSE} using the DM density field. For this visualization, we used the same persistence level (7$\sigma$) to extract the cosmic web for both gravity theories. It can be seen that this alternative gravity theory enhances more abundant filament structures. On bottom panels, the orange circles are drawn at one virial radius of their halo with $M_{\rm vir}>10^{13}\msun$, identified by \texttt{adaptaHOP}. For this zoomed region, we also represent the cosmic filaments in blue reconstructed by \texttt{DisPerSE} for the halo distribution using the CP$_{\rm max}$-halos calibration with parameters specified in Table~\ref{tab1}.}
\label{VisuGO}
\end{figure}

Measuring connectivity is non-trivial, since it depends on how filaments have been defined. Filament identification is dependent on the choice of tracers \citep[e.g.][]{Zakharova23}, the sampling of these tracers and the chosen tool and its parameterisation \citep[see e.g.][]{Libeskind18}.  Therefore in this paper, we try to adopt a definition of connectivity as objective as possible, and we focus on specifically measuring the connectivity of groups and clusters \citep[instead of measuring the connectivity of all peaks of the density field, as done in][]{2020MNRAS.491.4294K}, around which cosmic filaments are generally better constrained from observations, due to their high density contrast. Beyond factually checking the connectivity of the cosmic web under the Monge-Ampere gravity, our analysis aims at providing a framework for future comparison of the connectivity in alternative models and in large observational surveys. Furthermore, we aim to investigate whether our method can effectively exclude an alternative theory of gravity, particularly given its current developmental stage which suggests limited potential to replace Poisson gravity. Our method to measure connectivity is therefore driven by what is expected to be possible in future galaxy surveys. The paper is organized as follows. Section 2 provides a description of our data from cosmological simulations of Poisson and Monge-Ampère gravity and of how we measure the connectivity of massive halos. In Section 3, we describe our new metrics, based on the connectivity, designed to differentiate gravity theories. Moreover, these metrics hold the potential for extension to alternative DM theories. Section 4 summarizes our main results and suggests a benchmark for future studies.

\section{Data and methods}

\subsection{DM-only cosmological simulations}

The analysis presented in this paper uses snapshots of DM only simulations with Poisson gravity and Monge-Ampère gravity, starting from the same $\Lambda$CDM initial conditions using the N-GENIC code \citep{2012MNRAS.426.2046A,2005Natur.435..629S}. Our Poisson gravity simulation refers to classic $\Lambda$CDM cosmological simulations. These cosmological simulations from $z=49$ to $z=0$ were developed in a first work about Monge-Ampère gravity in a cosmological context ({see \cite{2024arXiv240407697L}). With a mass resolution of $m_{\mathrm{DM}}= 8 \times 10^{9} \msun$ and a uniform periodic-boundary cube of $L = 205$ Mpc/h $\sim 300$ Mpc on a side, these simulations enable us to study the clustering of DM at different scales, from galaxies to filaments. The cosmological parameters that we used are taken from \cite{2016A&A...594A..13P} results, with the matter density $\Omega_{\rm m} = 0.3089$, the cosmological constant $\Omega_{\rm \Lambda} = 0.6911$, the Hubble constant $H_0 = 67.74$ km s$^{-1}$ Mpc$^{-1}$ and $\sigma_8$ = 0.8159. The numerical implementation of this alternative gravity theory for a cosmological box using the latest algorithms based on optimal transport-based theory is detailed in \citep{2024arXiv240407697L}. Besides, the derivation of Monge-Ampère gravity from statistical physic principles is given in \cite{brenier:hal-01137528} and \cite{2024arXiv240407697L}. At each time step, the Monge-Ampère equation is solved using the optimal transport algorithm (fully detailed in \cite{refId0}). Then, gravitational forces are computed and momenta and positions are updated, as in standard N-body codes such as Gadget. Ultimately, the only difference from a classic N-body approach lies in the definition and calculation of the gravitational force. The large-scale structures have globally the same shape for both simulations, which was expected since Monge-Ampère theory can be considered as a small perturbation of Poisson gravity. As with Poisson gravity, MAG is a theory built without free parameters. This is why the two are identical in 1D. While one of its limitations is that it cannot be tuned, this also makes it a powerful framework for testing our approach. According to the power spectra  \citep[shown in Figure 9 of][]{2024arXiv240407697L}, at large scales, $\Lambda$CDM and this alternative gravity theory have the same trend since in low density regions far away from dense regions, the nonlinearities are negligable. However, on small scales, Monge-Ampère has much less power than $\Lambda$CDM. This behavior can also be observed through the percentage difference between Monge-Ampère and CDM at different scales shown in  Figure~\ref{DMnature}.

\begin{figure}[!t]
\centering
\includegraphics[width=\hsize]{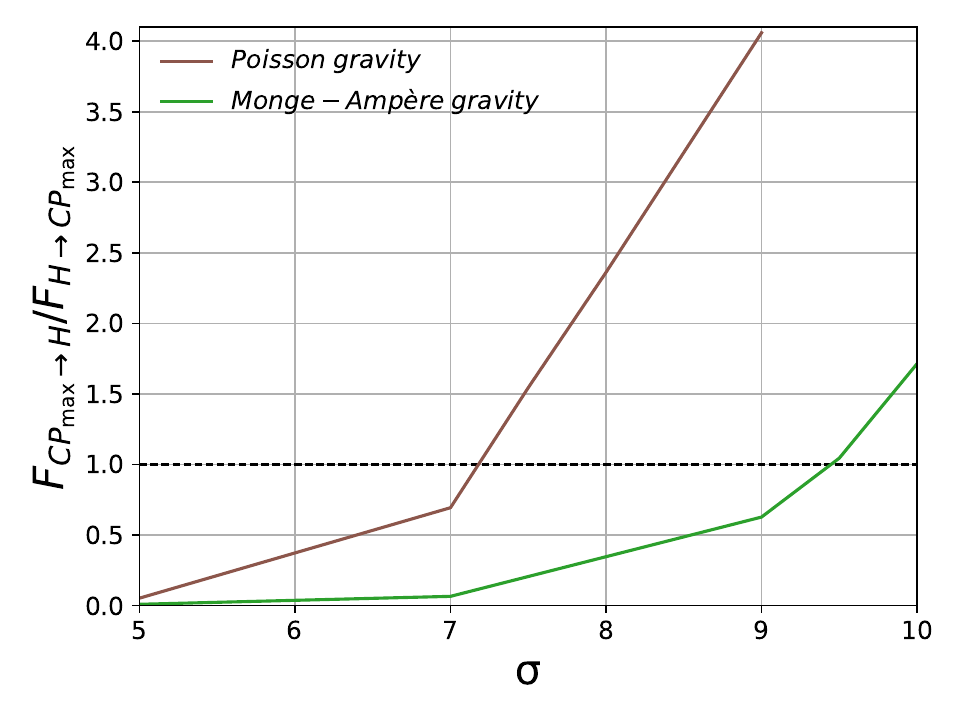}
\caption{Criterion to choose the persistence threshold for the cosmic skeleton based on DM density fields at $z=0$: ratio of the fraction of peaks (CP$_{\rm max}$) found within the virial radius (r$_{\rm vir}$) of any of the massive haloes to the fraction of groups or clusters hosting a CP$_{\rm max}$ inside their r$_{\rm vir}$ sphere, as function of the persistence parameter $\sigma$. We are seeking to determine for which persistence we are closest to unity.}
\label{Fract}
\end{figure}

\subsection{Extracting dark matter halos}

We used the \texttt{adaptaHOP} structure finder \citep{2004MNRAS.352..376A,2009A&A...506..647T} in the most massive submaxima mode (MSM) to identify the halos in our simulations. We used a local threshold of 178 times the average matter density, with the local density of individual particles calculated using the 20 nearest neighbours. The MSM method is designed to construct the halo structure tree (host halo along with its subhalos and nested subhalos), ensuring that the halo itself encompasses the most massive local maximum. We require a minimum of 100 particles per (sub)halo. Figure~\ref{HMF} shows the halo mass function in our simulations at $z=0$. The alternative theory of gravity strongly affects the halo formation over cosmic time. Monge-Ampère simulation exhibits a lack of DM bound structures, especially at high masses ($> 10^{13} \msun$). We emphasise that Monge-Ampère equation is invariant under a larger group of symmetries, i.e. SL(3) or unimodular affine symmetry, than the Poisson equation which is invariant only under SO(3) and does not support deformation or shearing transformations. The extra symmetry facilitates the formation of anisotropic structures such as filaments, but it also makes them more resistant to collapse, which reduces the formation of halos (see Figure~\ref{HMF}). This result can also be visually inspected on Figure~\ref{VisuGO}. Given the halo mass function induced by the Monge-Ampère gravity, it appears challenging to reconcile it with the observed cluster mass function. This further confirms that this theory is not viable, even if we have not resolved baryonic physics in our simulations with Monge-Ampère gravity. Indeed, the calculated halo mass function in Figure~\ref{HMF} is sufficient to exclude this gravity theory. Indeed, there is a difference of one to two orders of magnitude between the predictions of Poisson and Monge-Ampère. However, the observations with their error bars regarding the halo mass function are tightly constrained in our considered mass range and are in very good agreement with $\Lambda$CDM \citep{2022MNRAS.515.2138D}. On the contrary, our halo mass function in the Monge-Ampère gravity is clearly in disagreement with $\Lambda$CDM, considering the observation error bars. As mentioned before, our approach aims to find new metrics to constrain alternative gravity theories, as well as alternative DM theories that are consistent with certain observations, such as the stellar or cluster mass function.

\subsection{Extracting the cosmic web with \texttt{DisPerSE}}

We detect and extract the skeleton of the cosmic web using the publicly available algorithm Discrete Persistent Structure Extractor \citep[\texttt{DisPerSE},][]{2011MNRAS.414..350S,2011MNRAS.414..384S} applied to snapshots at redshift $z=$ 0, 0.5, 1, 1.5, 2 of both Poisson and Monge-Ampère simulations (see Figure~\ref{VisuGO}). This algorithm identifies the cosmic skeleton using the Discrete Morse Theory and the theory of persistence. In the case of DM-only cosmological simulations, it is possible to extract the cosmic web either from the density field traced by DM particles, or from catalogs of discrete objects such as DM halos. We opt for a dual approach, using these two cosmic web tracers, as leveraging the density field allows us to validate our simulations and methodology against simulation results from the literature. Then, extracting the cosmic skeleton based on halos enhances our alignment with observed data utilizing catalogs of galaxies.

\begin{figure}[!t]
\centering
\includegraphics[width=\hsize]{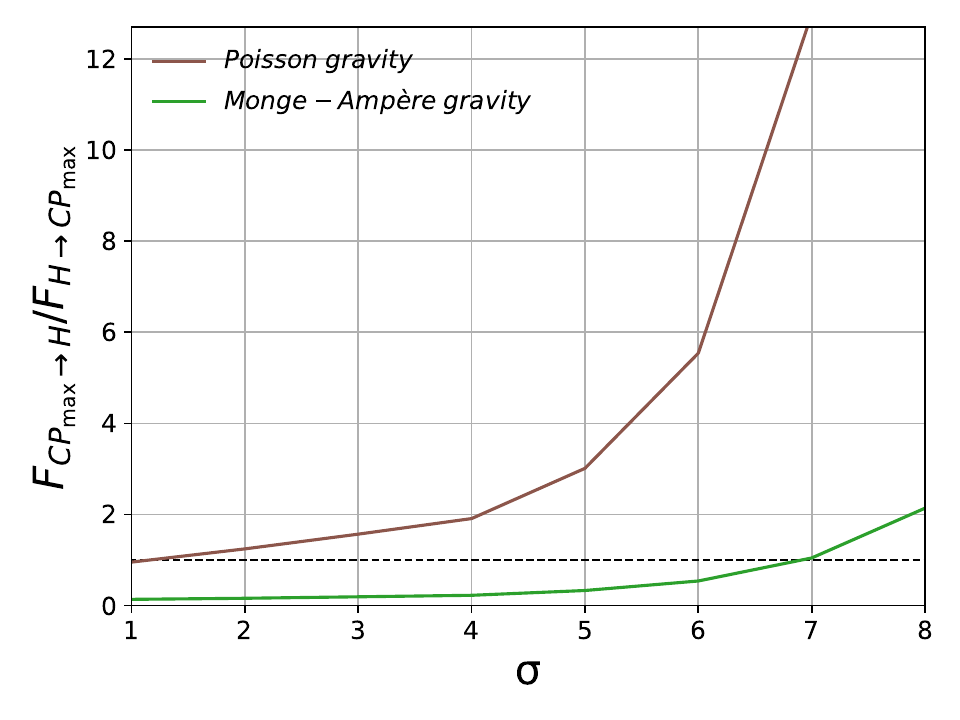}
\caption{Similar to Figure~\ref{Fract} expect that the cosmic skeleton is based on DM halo distributions at $z=0$.}
\label{FractHGD}
\end{figure}

\subsection{Calibration of \texttt{DisPerSE} \label{calib}}

Each filament is defined as a set of connected small segments linking peaks to saddle points. These endpoints of filaments are considered as critical points. In order to correctly extract filaments from our simulations, we need to assess the robustness of the derived skeleton by \texttt{DisPerSE}. One way to do that is to compare the positions of peaks (CP$_{\rm max}$) to those of the most massive halos such as galaxy groups and clusters ($> 10^{13} \msun$) identified by \texttt{adaptaHOP} for different persistence thresholds. The persistence is defined as the ratio of the density of the two critical points in the pair. This parameter, expressed in terms of numbers of $\sigma$, controls the significance level of the identified filaments with respect to the noise. Increasing the persistence threshold results in a decrease in the number of filaments as we are filtering low-persistence structures. This rigorous method of calibration was already been employed by \cite{2020A&A...641A.173G,2021A&A...649A.117G,2023A&A...671A.160G,2023arXiv230908659G}. During the skeleton construction, we also ensure that the filaments are not described by overlapping segments by using -breakdown option of \texttt{DisPerSE} in order to avoid double-counting filaments in the connectivity measurement. The results of the matching between CP$_{\rm max}$ and halos can be found in Figure~\ref{Fract} and ~\ref{FractHGD}. Instead of plotting the fraction of CP$_{\rm max}$ found within the virial radius (r$_{\rm vir}$) of any of the massive haloes, and conversely, the fraction of groups or clusters hosting a CP$_{\rm max}$ inside their r$_{\rm vir}$ spheres, we show the ratio of the two fractions, which means we are seeking to determine for which persistence we are closest to unity in order to minimize at $z=0$ both the number of halos without peak association and the number of peaks without halo association. The persistence thresholds chosen to compute the connectivity from skeleton extracted from the density field and halos are summarized in Table~\ref{tab1}. Cosmic skeletons at $z=$1, 2 were computed using the same \texttt{DisPerSE} parameterisation as for the $z=0$ case.

\begin{figure}[t!]
\centering
\includegraphics[width=\hsize]{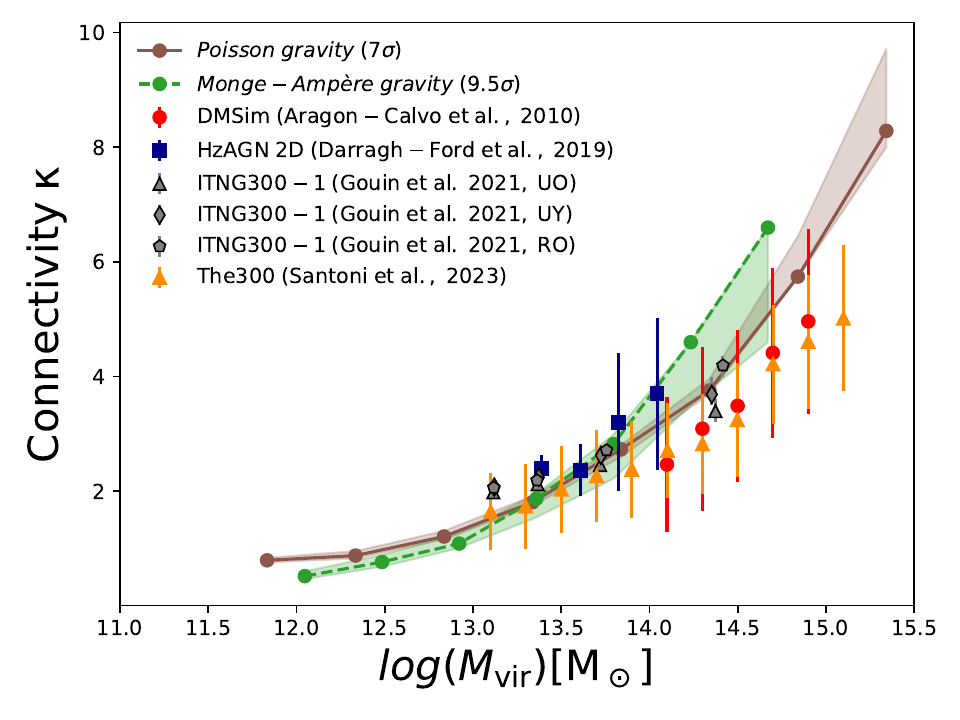}
\caption{Connectivity plotted against halo mass based on DM density distributions at $z=0$ with r$_{vir}$-2r$_{vir}$ band. Table~\ref{tab1} provides details on simulated data in the literature. Data from \cite{Gouin2021} are displayed for different dynamical states of galaxies.}
\label{Connect512BestM}
\end{figure}

\subsection{Connectivity $\kappa$}\label{DCon}

By design, we are mostly concerned here by the connectivity of galaxy groups and clusters given the resolution of our simulations and the next-generation surveys. One of the objectives of this paper is to investigate whether the connectivity can be a good metric to validate or invalidate alternative gravity theories such as Monge-Ampère gravity in a cosmological context and ultimately extend it to alternative DM theories such as WDM, SIDM, and FDM. The local number of filaments that are connected to galaxy groups and clusters in a sphere of 1.5 r$_{\rm vir}$ is the connectivity $\kappa$, similarly to \cite{2019MNRAS.489.5695D,Gouin2021,Sarron2019}. Given that the choice of a radius of 1.5 times the virial radius is arbitrary for calculating the mean connectivity, measurements within 1 and 2 times the virial radius are considered as error bars in all our figures.

Before assessing the impact of our alternative theory on the connectivity, we want to verify that our analysis of our $\Lambda$CDM DM-only simulations allows us to reproduce the results found in the literature. Table~\ref{tab1} compiles all key details about simulated and observed data from the literature, along with their analysis parameters regarding connectivity. Figure~\ref{Connect512BestM} attests to a good agreement between our Poisson simulation and state-of-the-art hydrodynamical cosmological simulations such as Horizon-AGN \citep{2014MNRAS.444.1453D}, Illustris TNG-300 \citep{2019ComAC...6....2N} and The Three Hundred \citep{2022MNRAS.514..977C}. Even if we obtain similar results to the literature, a rigorous calibration method based on the matching between CP$_{\rm max}$ and massive halos allows for setting a persistence threshold adapted to the simulation resolution. It would be interesting to see if the observed deviation at high mass ($>10^{14.5} \msun$) between our simulations is due to the inclusion of baryonic physics. In our study and on Figure~\ref{Connect512BestM}, we did not incorporate \cite{2023arXiv230908659G} data from the MTNG simulation. In their analysis, they chose to consider only halos with a connectivity of at least 2 \footnote{D. Galárraga-Espinosa, private communication}. This results in a significant deviation between our simulations and the MTNG simulation \citep{2023MNRAS.524.2539P} for masses less than $10^{14} \msun$, and this deviation becomes more pronounced as the mass decreases. Applying a similar criterion in our data, we observed this plateau behavior between $10^{12}$ and $10^{15}$ $\msun$. We also did not include \cite{2020MNRAS.491.4294K} data because they only looked at the connectivity of those galaxies which are located at the nodes of the cosmic web whereas in our case we measure the connectivity of all groups and clusters in the sample regardless the fact that they are sited on a node.

\begin{figure}[t!]
\centering
\includegraphics[width=\hsize]{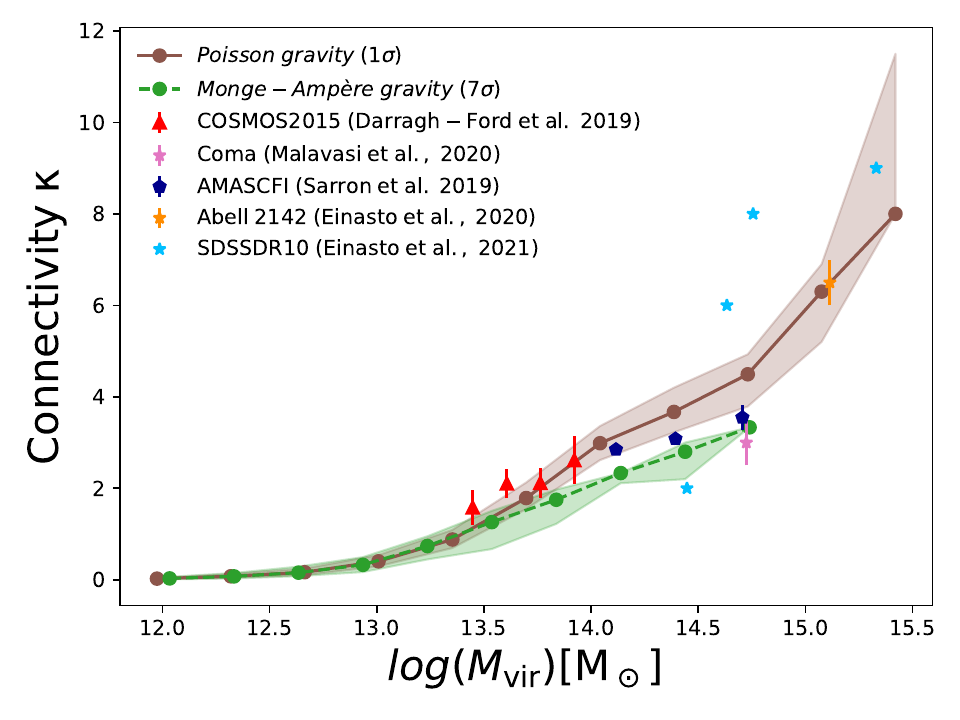}
\caption{Connectivity plotted against halo mass for the cosmic skeleton based on DM halo distributions at $z=0$ with r$_{vir}$-2r$_{vir}$ band. Table~\ref{tab1} provides details on observed data in from the literature.}
\label{Connect512BestMH}
\end{figure}

Figure~\ref{Connect512BestM} serves both to recall results from the literature and to highlight the importance of a rigorous calibration method when using filament finders such as DisPerSE. By adopting the same calibration approach, our CDM-only simulation produces results very similar to those of the TNG-300 simulation. We also note that the absence of baryons does not significantly impact connectivity in this mass range, as our results align with those of TNG-300 and Horizon-AGN 2D, both of which applied a nearly identical DisPerSE calibration method. This is consistent with the findings of \cite{2018MNRAS.473...68C} and particularly \cite{2023PhRvD.107b3514S}, who showed that even though halos lose more than 10$\%$ of their mass due to baryons, this mass transfer does not significantly shift the boundaries of DM structures. This is expected, as DM dominates the large-scale structure, while baryonic physics primarily affects small, highly non-linear scales such as galaxies and galaxy clusters. These non-linear baryonic effects have limited influence on the overall geometry of large-scale structures. Consistently, \citet{2011MNRAS.415..364V} (see also \citealt{2008ApJ...672...19R}) showed that baryons effects can be ignored on scales $k \geq 0.3 \; \mathrm{h/Mpc}$ when comparing hydrodynamic and DM-only runs. The disagreement with \cite{2010MNRAS.408.2163A} can be attributed to differences in connectivity measurements, as their study relied on a Multiscale Morphology Filter method, which appears to be less precise than DisPerSE. In contrast, the discrepancy with The Three Hundred simulations arises from their choice of gas as a tracer and the absence of a calibration method. It is also important to consider the potential impact of different seeds or box sizes in cosmological simulations. In CDM, the choice of the seed does not significantly affect the global connectivity statistics. While small halos, where initial fluctuations play a crucial role, could exhibit different connectivity, these halos are not resolved in our study due to resolution limitations. Additionally, the box size only constrains the range of halo masses for which connectivity can be measured (approximately $10^{12}$ and $10^{15}$ in our study). For instance, it was recently shown that changing the seed or the box size does not affect the filament length \citep{2024MNRAS.tmp.2584W}. Therefore, we conclude that the differences observed in Figure~\ref{Connect512BestM} are primarily due to the parameters used for computing connectivity rather than inherent variations between the simulations themselves. This makes connectivity a promising probe, provided that it is rigorously calibrated.

Figure~\ref{Connect512BestM} also compares the mean connectivity of Poisson and Monge-Ampère theories extracted from the DM particles. As the extra symmetry enhances the formation of filaments in Monge-Ampère gravity, it results that the mean connectivity is higher at high halo mass compared to Poisson gravity. We also stress that in both gravity theories, the connectivity increases with mass in agreement with previous studies. Besides, the trend of the mean connectivity approaching zero at low halo mass must be attributed to the lack of resolution in our simulations (see Figure~\ref{Connect512BestMH}). We emphasize that we take into account halos that have zero connectivity according to \texttt{DisPerSE} code. Figure~\ref{Connect512BestMH} contains the entire set of observed data that we have compiled from the literature (see details in Table~\ref{tab1}). In the case of the cosmic web extracted from DM halos, there is still a deviation between Poisson connectivity and that of Monge-Ampère at high halo mass (see Figure~\ref{VisuGO}). This behaviour persists because a cosmic web estimated from halos slightly underestimates the actual connectivity revealed by the DM density field (see Figure~\ref{DtoH}). The connectivity based on halos as a function of connectivity based on DM particles was computed by fitting the mean connectivities presented in the Figures~\ref{Connect512BestM} and ~\ref{Connect512BestMH} by excluding the points from the zero-slope region. We also examined whether the connectivity of the same halos identified in both simulations was correlated. This analysis, applied to 742 halos at $z=0$, revealed a positive correlation, as expected from Figure~\ref{Connect512BestMH}, but it remains very weak between halos formed in Poisson gravity and Monge-Ampère gravity. The choice to measure connectivity on halos rather than on the density field is driven by the ability to apply the same methods to observable galaxy data, as previously done by \cite{2019MNRAS.489.5695D, Gouin2021, 2020MNRAS.491.4294K, 2023arXiv230908659G}, and as currently implemented within the Euclid consortium. The simulation data concerning the connectivity used in our study of Poisson and Monge-Ampère models are available here \footnote{https://www.iap.fr/useriap/boldrini/data.html}.

\subsection{Evolution with redshift}

Future large surveys, like the one provided by the Euclid mission will allow to explore cosmic filaments around clusters at higher redshifts with more statistics, very useful especially at high mass where the difference between our two models is the most important, hence providing an extraordinary dataset to robustly test gravity models. This serves as motivation to explore the evolution of the mean connectivity as a function of redshift for both Poisson and Monge-Ampère simulations. In Figure~\ref{Connect512BestMHZ}, we found in our Poisson simulations that connectivity increases with redshift, as predicted by \cite{2023arXiv230908659G}. We remind that we extracted the high-z skeletons with the same persistence threshold as identified at $z=0$. Starting from Figure~\ref{Connect512BestMHZ}, interpreting the behavior of connectivity with respect to redshift is challenging. However, by calculating the mean connectivity, we observed that it remains almost constant between $z=1$ and $z=2$, after which it increases until $z=0$, similar to the Poisson model. This strongly encourages us to evaluate the connectivity as a function of redshift, as the differences between our gravity models will manifest not only at $z=0$ but also over time. Indeed, we remember that the Monge-Ampère gravity influences the formation of structures such as filaments and halos.

\begin{figure}[t!]
\centering
\includegraphics[width=\hsize]{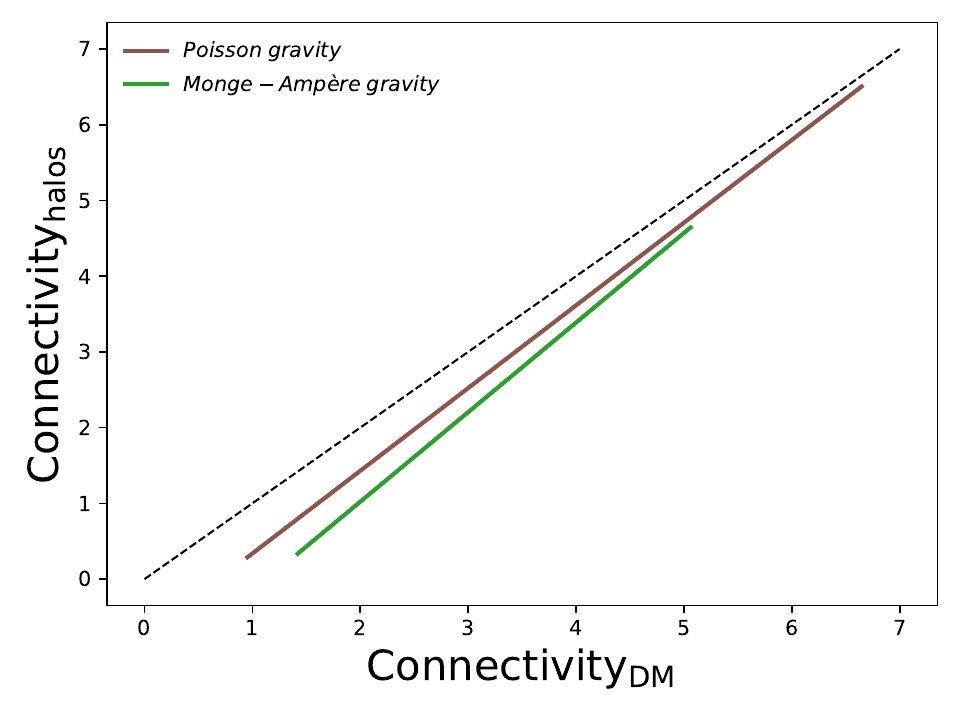}
\caption{Connectivity based on halos as function of connectivity based on DM particles. Both connectivities were computed as functions of the halo virial mass in Figure~\ref{Connect512BestMH} and ~\ref{Connect512BestM}. A cosmic web estimated from halos slightly underestimates the actual connectivity revealed by the DM density field. This effect is stronger for our alternative theory of gravity. Despite the small difference between the connectivities, this also strengthens the idea that it is preferable to extract the skeleton from the halos if one wants to follow an observational approach.}
\label{DtoH}
\end{figure}

\section{Results}

In light of Figure~\ref{Connect512BestMH}, the question that now interests us is how we can utilize these observed connectivity data to distinguish between our gravity theories and, subsequently, alternative DM theories. As depicted by this Figure, it is not possible to precisely distinguish between our two models with the observed data when considering the connectivity as a function of the virial mass at fixed redshift. However, the redshift evolution of the signal is supposed to carry information on the overall growth of structures governed by gravity.

\subsection{Redshift evolution of the slope and offset of the connectivity}

The two metrics considered to evaluate our theories are the connectivity slope and offset with respect to redshift (see Figures~\ref{SlopeK} and ~\ref{SlopeM}). They are calculated by fitting the mean connectivity as a function of the virial mass for each redshift. In order to mitigate the contamination of numerical noise due to the lack of resolution in our simulations, we exclude points from the zero-slope region for our fits. We only keep those with a slope greater than 0.1. The offset corresponds to the y-intercept of the fitted function for a mass of $M_{\rm vir}=10^{14}$ M$_{\odot}$. This arbitrary mass marks approximately the center of the region where there is the highest density of observable data (see Figure~\ref{Connect512BestMH}). Relying on the slope and offset of connectivity as robust metrics is motivated by their ease of tracking, particularly when evaluating connectivity across different redshifts. In order to compare to observations, we have calculated our two metrics for the COSMOS2015 \citep{2016ApJS..224...24L} and CFHTLS data \citep{2018A&A...613A..67S}. We have also chosen to include a data point which takes into account all the observable data reported in Table~\ref{tab1}. To do this, we have assigned a weight to the data corresponding to the number of objects identified in the mass bin as reported by the associated publications. In this calculation, we have not included connectivity measurements of the four richest clusters of the Corona Borealis supercluster \citep{Einasto21} as they do not have error bars. Regarding the Poisson theory, we observe that the slope and the offset are decreasing and increasing functions with respect to redshift, respectively (see Figures~\ref{SlopeK} and ~\ref{SlopeM}). However, it is more challenging to identify a linear behavior as in $\Lambda$CDM, even though the slope appears to be almost constant with respect to redshift for Monge-Ampère gravity.

\begin{figure}[t!]
\centering
\includegraphics[width=\hsize]{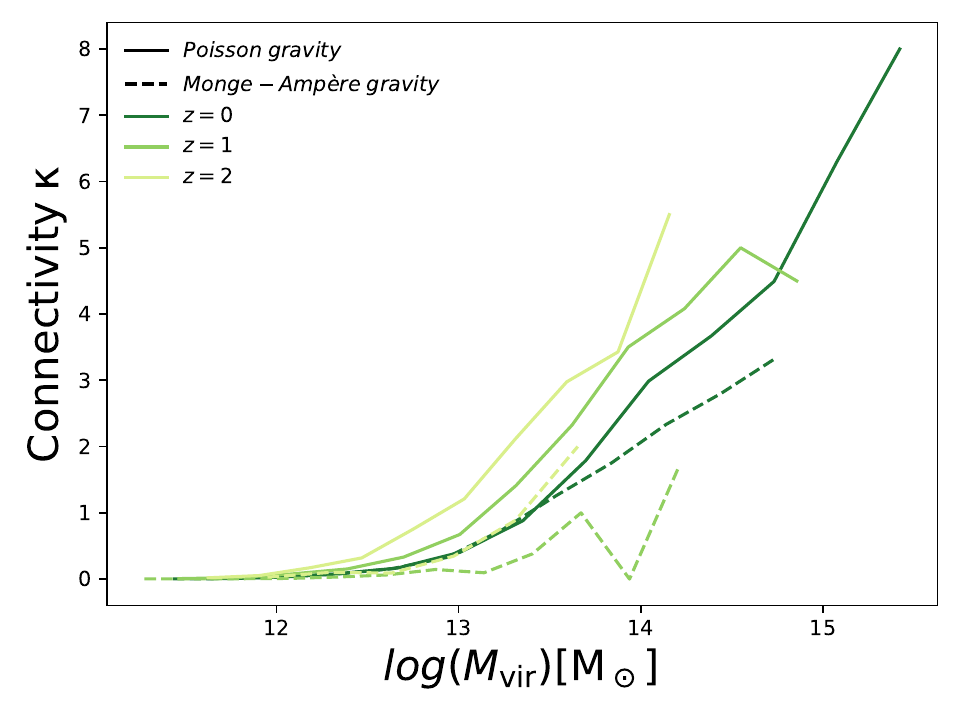}
\caption{Connectivity plotted against halo mass for the cosmic skeleton based on halo distributions at $z=0,1$ and 2.}
\label{Connect512BestMHZ}
\end{figure}

According to Figure~\ref{SlopeK}, the slope of connectivity clearly distinguishes our gravity theories up to $z=1$. Although the considered data do not allow us to favor one gravity theory over the other, it appears that the overall slope, taking into account all the literature data, aligns with the Poisson theory. This result is strengthened by the fact that all our data points for the connectivity offset are in very good agreement with the $\Lambda$CDM gravity theory. It is worth noting that this metric significantly differentiates our two gravity theories between $z=0$ and $z=2$. Our results highlight that the data from the next generation surveys, probing a larger range of redshift than nowadays observations with increased statistics, will be crucial for robustly validating or invalidating alternative theories through these metrics. The data concerning the connectivity slope and offset used in our study are also available here \footnote{https://www.iap.fr/useriap/boldrini/data.html}.

\begin{figure}[t!]
\centering
\includegraphics[width=\hsize]{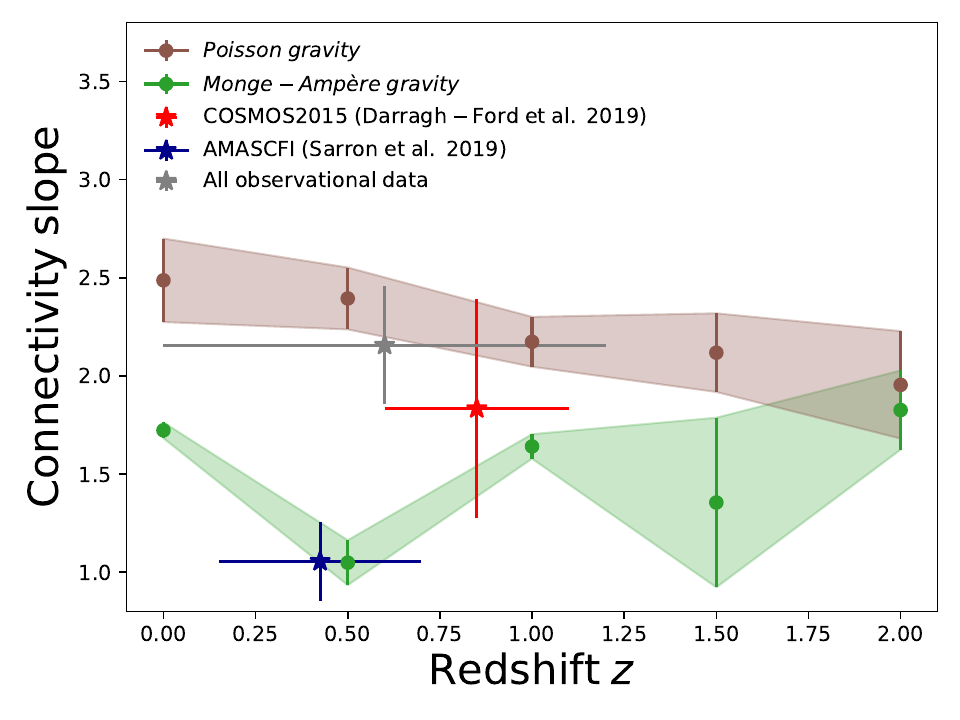}
\caption{Connectivity slope as a function of redshift. For simulations, measurements within 1 and 2 times the virial radius are still considered as error bars. The gray point corresponds to the slope that takes into account all the observable data from Table~\ref{tab1}.}
\label{SlopeK}
\end{figure}

\subsection{Alternative DM theories and cosmic web}

A question that now arises concerns the viability of our approach not only for gravity theories affecting the dynamics of matter, but for alternative theories of DM whose intrinsic properties are modified. Intuitively, one would expect our new metrics to work within the framework of DM theories that affect the cosmic web and, consequently, the connectivity of filaments. Indeed, we have highlighted that modifying the gravity in Monge-Ampère theory notably impacts large-scale dynamics, and this modification leaves an imprint on the connectivity. Are there alternative DM theories with exotic properties that affect the cosmic skeleton? A wide range of alternative DM models has been proposed over the last decades. Two main classes of alternative DM models have been simulated on large scales: warm DM  \citep[WDM,][]{2001ApJ...556...93B} and fuzzy DM  \citep[FDM,][]{2000PhRvL..85.1158H}. These theories have gained attention as alternatives to the standard CDM model because they offer potential solutions to some of CDM puzzles, particularly on small scales \citep{2017ARA&A..55..343B,2021Galax..10....5B}. That's why there are few studies, especially on large scales, concerning the cosmic web and its properties.

Since particles in WDM and FDM models have different intrinsic properties from the CDM particle candidates, the effect of these particles on structure formation and evolution is expected to be qualitatively different on both small and large scales and particularly on the overdense structures that connect halos \citep{2015arXiv150603789P,2020MNRAS.494.2027M,2023MNRAS.524.4256M}. In WDM cosmology, it has been demonstrated that the presence of less substructures is associated with warmer DM. Specifically, as the particle velocity dispersion increases, the filaments become larger, and their preservation extends over a longer period. For WDM models, structure formation follows a complex scenario in which both top-down and bottom-up processes play a role \citep{2015arXiv150603789P}. Due to this hybrid structure formation, we expect that this difference will be evident in the connectivity across redshifts. Like WDM, FDM filaments do not fragment into halos like those of CDM \citep{2020MNRAS.494.2027M,2023MNRAS.524.4256M}. Over time, CDM filaments can fragment and collapse to form gravitationally bound structures known as DM halos, which are seeds for the formation of galaxies and galaxy clusters. This is why WDM and FDM simulations appear similarly filamentary \citep{2020MNRAS.494.2027M}. Furthermore, these alternative models can also induce a delay in the formation of structures. In CDM scenario, the first structures begin to form well before reaching $z = 9$. In contrast, in a WDM universe, the first collapse may be delayed until $z = 4$ \citep{2015arXiv150603789P}. 

Figure~\ref{DMnature} presents a comparison of power spectra normalized to CDM for the Monge-Ampère model and two alternative DM (ADM) theories as a function of the spatial scale of structures at $z=3$ and 0. Given that the deviations in the power spectrum of the Monge-Ampère model from CDM are sufficient to induce visible differences in halo connectivity, we use this to infer the impact of FDM and WDM spectra on connectivity by comparing their respective deviations from CDM. By visual inspection, the WDM spectrum at $z=0$ from \cite{2012MNRAS.421...50V} also shows a suppression of DM clustering, which is more pronounced at small scales than in the Monge-Ampère model. Unlike WDM, where the cutoff is gradual, FDM at $z=3$ from \cite{2023MNRAS.524.4256M} exhibits a sharp cutoff at small scales due to quantum pressure effects. Therefore, we expect these alternative DM theories to disrupt the connectivity of their halos, as they exhibit cutoffs similar to the gravity theory investigated here but at different scales (see Figure~\ref{DMnature}). Indeed, in WDM and FDM scenarios, a suppression of the number of halos is expected compared to the CDM model, particularly for $M<=10^{9}M_\odot$, which will disrupt the connectivity of more massive halos because connectivity is simply the count, within a given radius, of the number of filaments traced by the surrounding halos. All these physical arguments from the literature motivate us to investigate the properties of filaments in these alternative theories, under a benchmark framework that we propose. We emphasize that the expected impact on connectivity remains a hypothesis to be tested, which reinforces the need for further studies.

\begin{figure}[t!]
\centering
\includegraphics[width=\hsize]{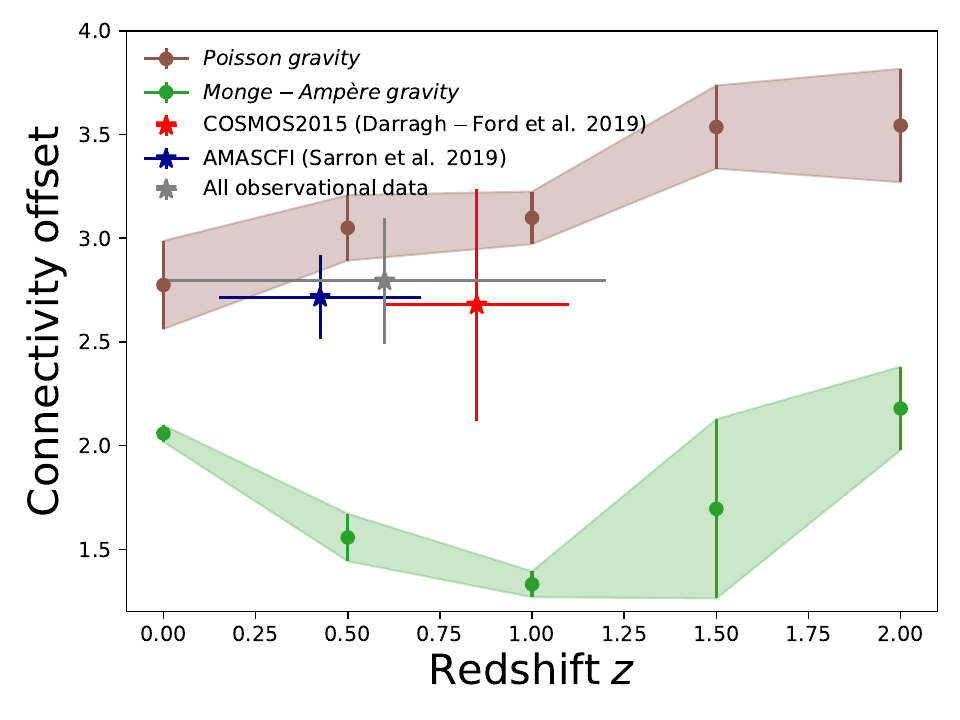}
\caption{Connectivity offset as a function of redshift for $M_{\rm vir}=10^{14}$ M$_{\odot}$. For simulations, measurements within 1 and 2 times the virial radius are still considered as error bars. The gray point corresponds to the offset that takes into account all the observable data from Table~\ref{tab1}.}
\label{SlopeM}
\end{figure}

\subsection{Forecasts for future surveys}
While Fig.~\ref{SlopeK} and Fig.~\ref{SlopeM} show that observations are in overall better agreement with the Poisson gravity model, error bars remain large and some data points at specific redshift (e.g. the AMASCFI dataset at $z\sim 0.5$) are clearly discrepant with respect to the bulk of the observations. This indicates both the need for larger group and cluster samples around which to measure connectivity to beat down the statistical uncertainties, and a consistent measurement of connectivity across redshift ranges. The next-generation of photometry surveys are particularly well suited to address these issues. For example, the European Space Agency's \textit{Euclid} mission \citep{euclid} will provide, over the entire course of the mission, several thousands of~clusters  \citep{Adam19euclidcluster} over $\sim 14000 {\rm deg}^2$ to $z\sim 2$ and down to $10^{13.5}M_\odot$. We note that the accuracy of photometric redshifts might be too limited in the \textit{Euclid} Wide Survey \citep{EWS} especially above $z=1$ to reconstruct well enough the distribution of cosmic filaments around clusters. However deeper photometric data will be available in the several tens of degrees of the \textit{Euclid} Deep survey, along with exquisite complimentary ancillary optical and infrared photometry from HSC and Spitzer as part of the Hawaii Two-0 Survey \citep{H20}, which should enable a photometric redshift accuracy of the same order of the one obtained on the COSMOS field \citep{Weaver2022}. In this case, we expect to multiply by  $\sim 20$ our number of detections with respect to the data reported from \cite{2019MNRAS.489.5695D} while making a consistent extraction across the redshift range to $0<z<2$, which will significantly improve the observational constraints on connectivity. Besides \textit{Euclid}, other upcoming surveys which have identified cosmic web mapping as one of their focus science cases could be used, including spectroscopic surveys like e.g. 4MOST \citep{dejong2012} and PFS \citep{Takada2014} or other photometric surveys like LSST \citep{lsst} or WFIRST \citep{WFIRST}.

\begin{figure}[t!]
\centering
\includegraphics[width=\hsize]{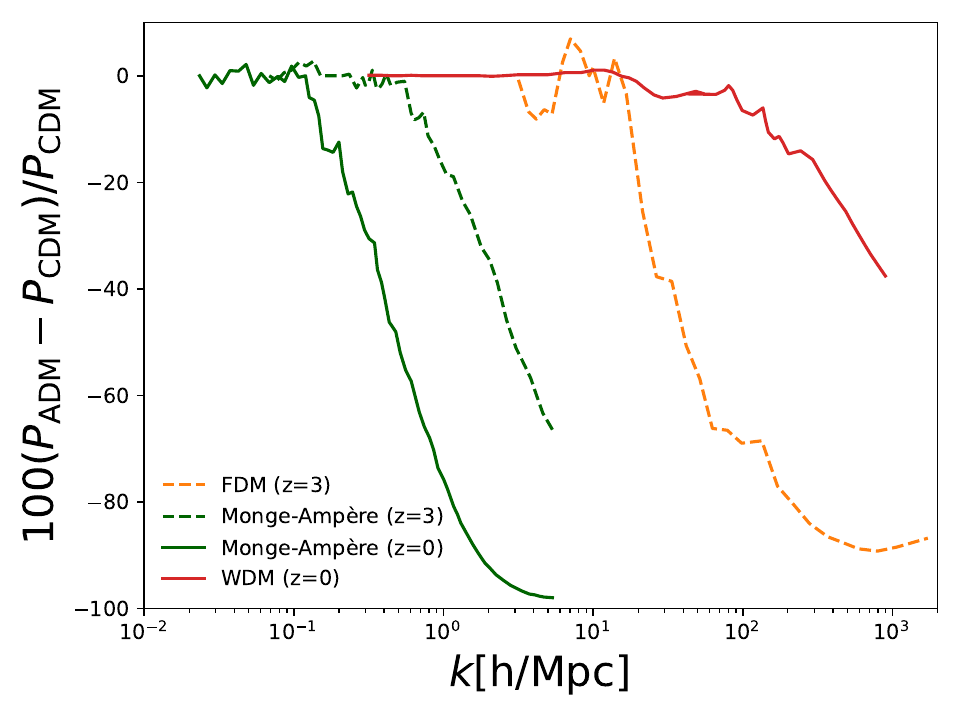}
\caption{Percentage difference between alternative DM theories (ADM) and CDM power spectra. The power spectra data for FDM and WDM are taken from \cite{2023MNRAS.524.4256M} and \cite{2012MNRAS.421...50V} with a DM mass parameter of 2.5 $\times10^{-22}$ eV and 2 keV at $z=3$ and $z=0$, respectively. The cutoffs observed in WDM and FDM, similar to those in the Monge-Ampère model but occurring at different scales, are expected to induce differences in halo connectivity compared to CDM.}
\label{DMnature}
\end{figure}

We note that we have not explicitly explored the effect associated to photometric redshift uncertainties and projection, which might impact the extraction of cosmic filaments, hence the connectivity measurements. In addition, different methods of cluster selections suffer from different biases, depending on if they are based either on galaxies probing primarily the stellar mass, or on X-Ray sensitive to the warm gas content, or on lensing measuring the total mass. The selection and cluster mass measurements methods will potentially create an additional source of systematics. We have not quantified these biases in this paper since the measurements presented here were not done in the context of forecasting a particular survey, and the observational data that we reported come both from spectroscopic and photometric samples, which make difficult a thorough analysis of biases. We note however that the good agreement between most of the surveys presented here suggest that observational uncertainties and systematics are sub-dominant with respect to the impact of the considered gravity model. However, a more careful end-to-end analysis of all systematic biases affecting specific surveys is necessary and will be presented in a future analysis.

\section{Conclusions}

In this work, we have explored the properties of the cosmic web in an alternative theory of gravity, the Monge-Ampère gravity. We used DM-only cosmological simulations of Monge-ampère and Poisson gravity ($\Lambda$CDM) models at redshifts $z=$ 0, 0.5, 1, 1.5, 2. Following an observational approach, we have extracted the cosmic skeleton using the \texttt{DisPerSE} filament finder by employing the calibration method suggested by \cite{2020A&A...641A.173G}. We then focused on the connectivity of groups and clusters in both theories in order to provide a framework for future comparison of its properties in alternative models and in large observational surveys. We have demonstrated that distinctions between our gravity theories begin to manifest in the measurement of the connectivity of high-mass halos ($M_{\rm vir}>10^{14}$ M$_{\odot}$). To precisely differentiate our theories, the redshift evolution of the slope and offset of the connectivity can be used to retrieve information on the overall growth of structures, which differs in our two theories.
Following our study of connectivity for different gravity theories, we propose a benchmark for a more effective comparison between simulations and observations:
\begin{itemize}
    \item Creating mock data, including noise and instrumental properties, to replicate the survey data based on simulations 
    \item Extracting cosmic filaments from the mock data and observations with the \texttt{DisPerSE} code
    \item Calibrating independently the persistence threshold of \texttt{DisPerSE} for observed and simulated data by using CP$_{\rm max}$-halos method
    \item Computing specific metrics to compare synthetic and observed data.
\end{itemize}
This approach enables unbiased comparisons between the data and theoretical predictions, free from the influence of the theories themselves. This benchmark will also enable us to make the most of the new metrics identified in our study: the slope and offset of connectivity across redshift. We have demonstrated that these metrics could serve as reliable probes to validate or invalidate gravity theories such as the Monge-Ampère gravity. Data from next-generation surveys such as Euclid will confirm the success of our approach for alternative models in the mass regime of groups and clusters. As suggested by our results, there is a clear need for larger samples of groups and clusters to measure observed connectivity. This is essential to reduce statistical uncertainties and ensure a consistent measurement across redshift ranges. Finally, we discussed how our approach can be extended to alternative DM theories such as warm or fuzzy DM.

\section{Data Availability}

The code, which computes the filament connectivity at a given halo mass and then the connectivty slope and offset, is publicly available at \href{https://github.com/Blackholan}{\textcolor{blue}{https://github.com/Blackholan}}. The data results shown in this paper can be found at \href{https://www.iap.fr/useriap/boldrini/data.html
}{\textcolor{blue}{https://www.iap.fr/useriap/boldrini/data.html}}.

\begin{acknowledgements}

The first part of P. Boldrini's work was carried out as part of a post-doc (funded by Inria Nancy and Sorbonne University), co-supervised by R. Mohayaee and B. Lévy, and in collaboration with Y. Brenier. The study of connectivity as a possible cosmological probe, which is the subject of this paper, was conducted outside of and after this post-doc. This work uses data from the Monge-Ampère gravitational numerical simulation \citep{2024arXiv240407697L}. We thank Céline Gouin and Daniela Galárraga-Espinosa for valuable discussions. P.B. also thanks Eduardo Vitral and Denis Werth for their help concerning the data visualisations. This work has made use of the Infinity Cluster hosted by the Institut d’Astrophysique de Paris.

\end{acknowledgements}

\bibliography{src}

\end{document}